 \documentclass[final,5p,times,twocolumn]{elsarticle}

\usepackage{amssymb}
\usepackage{amsmath}

\usepackage[ruled,vlined,linesnumbered]{algorithm2e}
\usepackage{threeparttable}
\usepackage{booktabs}
\biboptions{sort&compress}
\usepackage[pagebackref=false,breaklinks=true,letterpaper=true,colorlinks,bookmarks=false]{hyperref}

\usepackage[table]{xcolor}
\usepackage{booktabs}
\usepackage{threeparttable}
\usepackage{multirow}
\usepackage{arydshln}

\begin{document}

\begin{frontmatter}

\title{Discourse-Aware Scientific Paper Recommendation via QA-Style Summarization and Multi-Level Contrastive Learning}

\author[label1,label2,label3]{Shenghua Wang}
\author[label2,label3]{Zhen Yin\corref{cor1}}
\affiliation[label1]{organization={Institute of Communication Studies, Communication University of China},
             city={Beijing},
             postcode={100024}, 
            country={China}}
            
\affiliation[label2]{organization={Beijing Renhe Information Technology Co., Ltd.},
            city={Beijing},
            postcode={100096}, 
            country={China}}
            
\affiliation[label3]{organization={Key Laboratory of Digital Publishing and Total Process Management of Scientific and Technical Journals},
            city={Beijing},
            postcode={100083}, 
            country={China}}

\cortext[cor1]{Corresponding author}

\begin{abstract}
The rapid growth of open-access (OA) publications has intensified the challenge of identifying relevant scientific papers. Due to privacy constraints and limited access to user interaction data, recent efforts have shifted toward content-based recommendation, which relies solely on textual information. However, existing models typically treat papers as unstructured text, neglecting their discourse organization and thereby limiting semantic completeness and interpretability.
To address these limitations, we propose OMRC-MR, a hierarchical framework that integrates QA-style OMRC (\textit{Objective}, \textit{Method}, \textit{Result}, \textit{Conclusion}) summarization, multi-level contrastive learning, and structure-aware re-ranking for scholarly recommendation. The QA-style summarization module converts raw papers into structured and discourse-consistent representations, while multi-level contrastive objectives align semantic representations across metadata, section, and document levels. The final re-ranking stage further refines retrieval precision through contextual similarity calibration.
Experiments on DBLP, S2ORC, and the newly constructed Sci-OMRC dataset demonstrate that OMRC-MR consistently surpasses state-of-the-art baselines, achieving up to 7.2\% and 3.8\% improvements in Precision@10 and Recall@10, respectively. Additional evaluations confirm that QA-style summarization produces more coherent and factually complete representations. Overall, OMRC-MR provides a unified and interpretable content-based paradigm for scientific paper recommendation, advancing trustworthy and privacy-aware scholarly information retrieval.
\end{abstract}


\end{frontmatter}




\section{Introduction}
\label{introduction}

The exponential growth of scientific publications across disciplines has significantly increased the difficulty for researchers to efficiently locate relevant and high-quality literature \cite{altmami2022automatic,haddaway2022citationchaser}. Although traditional search engines and citation-based systems provide partial solutions, they often fail to capture nuanced thematic and methodological connections between papers, especially across domains \cite{haruna2018citation,kanwal2024research,pinedo2024arzigo}. Consequently, scholarly paper recommendation has emerged as a key functionality for digital libraries, research platforms, and academic discovery systems, aiming to mitigate information overload and enhance knowledge discovery.

Conventional recommendation approaches can be categorized into collaborative filtering, citation-based, and content-based methods. Collaborative filtering predicts user preferences based on past interactions \cite{wang2020collaborative,chen2024kgcf,g2024comprehensive}, but it suffers from cold-start issues and data sparsity. Citation-based systems leverage citation graphs to identify related literature, yet they rely on complete and up-to-date citation data, which is often unavailable for newly published or domain-specific documents. Content-based methods, on the other hand, recommend articles by analyzing intrinsic textual features. Traditional approaches such as TF-IDF and BM25 \cite{marwah2020term} capture lexical similarities, while more advanced models incorporate topic modeling \cite{chauhan2021topic,wu2024survey} or neural embeddings \cite{li2024graph}. Although pre-trained language models (PLMs) like BERT, SciBERT, and SPECTER \cite{darraz2025integrated,dinh2024enhancing,cohan2020specter} have significantly improved semantic representations, they often treat scientific texts as flat sequences and overlook the internal rhetorical structures inherent in academic writing. Furthermore, these models typically operate under monolingual assumptions and lack mechanisms for multilingual alignment.

Recent efforts have introduced large language models (LLMs) into scholarly recommendation, demonstrating impressive capabilities in zero-shot retrieval and user profile generation \cite{he2023large,bao2023tallrec}. However, LLM-based approaches are computationally expensive, difficult to scale, and often lack interpretability or grounding in explicit evidence. More importantly, most existing models fail to differentiate rhetorical components such as objectives, methods, and results, which are essential for identifying methodological relevance between papers. This limitation leads to reduced precision in recommendations, particularly in cross-disciplinary or structurally diverse scientific corpora.

In addition to these technical challenges, practical deployment environments impose further constraints \cite{zaguir2024challenges,vethachalam2024cloud}. With the widespread adoption of open-access publishing and the enforcement of stricter privacy regulations, most scientific platforms no longer collect or disclose detailed user interaction data. Citation information is frequently delayed or missing for recent, non-English, or underrepresented publications. These limitations underscore the growing need for recommendation systems that rely solely on document content, while still achieving semantic accuracy, structural awareness, and adaptability to multilingual scenarios. In this context, content-driven recommendation emerges as a robust and privacy-preserving solution for scalable scientific discovery, especially in the absence of user behavior logs or citation graphs.

\begin{figure}[t]
    \centering
    \includegraphics[width=1\linewidth]{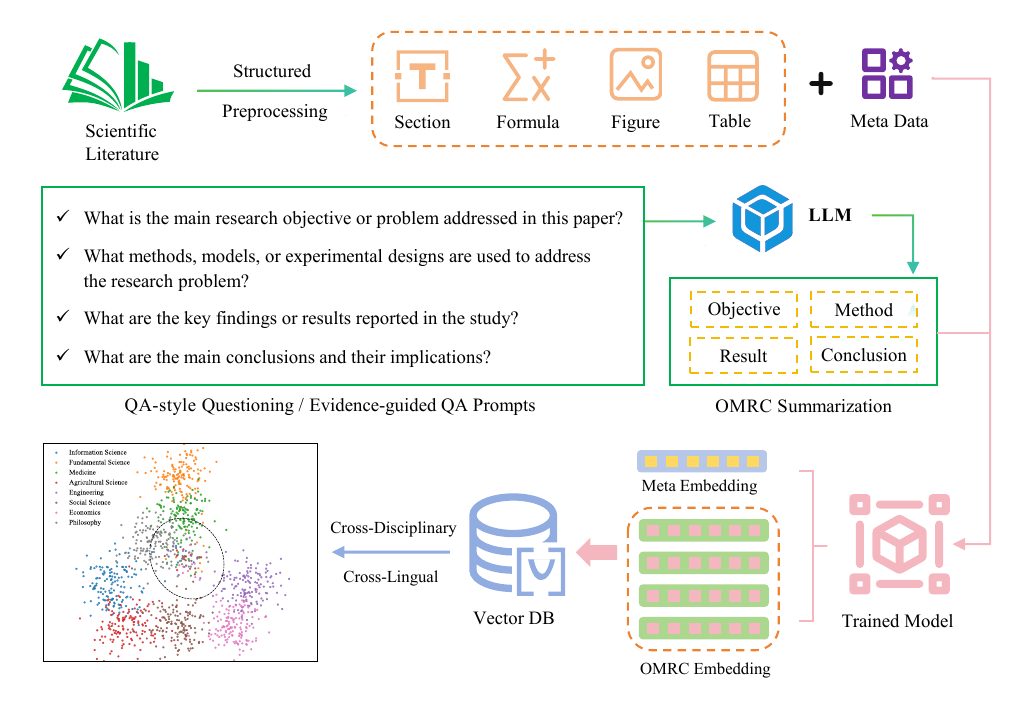}
\vspace{-.5cm}
\caption{
Conceptual overview of the proposed QA-style OMRC framework for cross-disciplinary and multilingual scholarly recommendation.}
\label{fig:Figure1}
\end{figure}

To address these challenges, we propose a unified framework named OMRC-MR, which integrates QA-style OMRC (\textit{Objective}, \textit{Method}, \textit{Result}, \textit{Conclusion}) evidence-constrained summarization, multi-level contrastive learning, and role-aware retrieval for scholarly paper recommendation. Each document is first transformed into structured OMRC summaries grounded in explicit textual evidence and aligned with metadata such as titles, abstracts, and keywords. These complementary textual views are then encoded using a multilingual pre-trained model and projected through role-specific semantic heads to preserve distinctions across rhetorical functions. Next, a multi-level contrastive learning objective is optimized to enforce metadata–summary alignment, role-level discrimination, and cross-lingual consistency. Finally, the resulting document embeddings support a hierarchical retrieval and re-ranking strategy, in which coarse-grained metadata similarity is refined by fine-grained role-aware similarity, yielding interpretable and cross-disciplinary recommendations. The overall architecture of OMRC-MR is illustrated in Fig.~\ref{fig:Figure1}.

The main contributions of this work are as follows:
\begin{itemize}
    \item[\textcolor{black}{$\bullet$}] We propose OMRC-MR, a hierarchical content-based recommendation framework that integrates QA-style OMRC summarization, multi-level contrastive learning, and structure-aware re-ranking to capture the discourse structure of scientific papers and improve semantic representation and retrieval precision.
    \item[\textcolor{black}{$\bullet$}] We construct Sci-OMRC, a large-scale dataset comprising English and Chinese scientific papers from multiple disciplines, with structured full-text elements such as sections, tables, figures, and formulas, supporting discourse-level modeling and evaluation.
    \item[\textcolor{black}{$\bullet$}] Extensive experiments on DBLP, S2ORC, and Sci-OMRC demonstrate that OMRC-MR consistently outperforms representative baselines. Further evaluations on automated summarization quality and component-wise ablation confirm that QA-style summarization, contrastive optimization, and re-ranking collectively contribute to the framework’s accuracy, interpretability, and robustness.
\end{itemize}

\section{Related Work}
\label{relatedwork}

\subsection{Scholarly paper recommendation}

The task of scholarly paper recommendation has been extensively studied to address the challenge of information overload faced by researchers. Early systems were predominantly based on collaborative filtering and citation graph analysis \cite{ko2022survey,zhang2023scholarly}, which exploited co-reading patterns and citation propagation to infer relatedness. With the introduction of content-based methods, textual features such as titles, abstracts, and keywords were incorporated into recommendation pipelines. Topic modeling approaches like Latent Dirichlet Allocation \cite{chauhan2021topic} enabled probabilistic representations of thematic structures, while neural embedding methods \cite{zhu2021recommending} captured semantic similarity more effectively. The emergence of PLMs such as BERT, SciBERT, and SPECTER \cite{beltagy2019scibert,cohan2020specter} further improved semantic-rich document embeddings, setting new benchmarks for paper recommendation. Recently, LLMs have also been explored in this field, either as zero-shot recommenders or as generators of structured profiles for scholarly articles \cite{zhu2024collaborative,zhang2025recommendation}, demonstrating their potential in capturing complex cross-document relations.

Despite these advances, several limitations persist. Existing PLM- and LLM-based approaches often assume monolingual or domain-specific corpora, restricting their effectiveness in cross-disciplinary and multilingual settings. Moreover, LLM-based recommenders, while flexible, suffer from high computational cost, lack of scalability, and limited controllability in evidence grounding. Current methods also rely heavily on flat textual inputs such as abstracts or titles, overlooking the internal heterogeneity of scientific papers, including structured roles such as objectives, methods, and results. This simplification leads to insufficient modeling of fine-grained research aspects, which are essential for distinguishing semantically related yet methodologically divergent works. Furthermore, in practical deployment scenarios, the availability of user interaction data has diminished significantly due to the growing adoption of open-access publishing and increasingly strict privacy regulations. Combined with the delayed or missing citation links for newly published or underrepresented papers, these constraints further highlight the need for content-based recommendation systems that rely solely on document content while maintaining semantic fidelity, structural awareness, and multilingual adaptability.

\subsection{Structured summarization of scientific literature}

Abstracts have long been regarded as the primary source of semantic representation for scholarly articles. Traditional summarization methods can be divided into extractive approaches, which select salient sentences from the text \cite{nallapati2017summarunner,azam2025current}, and abstractive approaches, which generate novel sentences to produce concise summaries \cite{gupta2019abstractive,giarelis2023abstractive}. In the context of scientific literature, efforts have been made to design domain-specific summarization systems, with pre-trained models like SciBERT or PEGASUS \cite{alsultan2025pegasus,feng2025enhancing} providing stronger language generation capabilities. Moreover, the IMRaD (Introduction, Methods, Results, and Discussion) structure has become a widely adopted template for organizing summaries, enabling better alignment with the rhetorical functions of research articles.

However, existing summarization methods exhibit several shortcomings for scholarly recommendation. Most approaches remain monolingual and rarely consider multilingual corpora, which limits their applicability in global scientific communication. In addition, while IMRaD-based summarization provides a structured view, it often lacks explicit evidence constraints that ensure factual consistency and verifiability. Furthermore, the alignment between metadata (e.g., \textit{title}, \textit{abstract}, \textit{keywords}) and structured summaries has not been adequately exploited, resulting in information redundancy and inconsistencies across different textual views. These gaps motivate the use of evidence-constrained OMRC (\textit{Objective}, \textit{Method}, \textit{Result}, \textit{Conclusion}) summarization in our framework, which enhances factual grounding and enables multi-perspective alignment for cross-disciplinary and cross-lingual recommendation.

\subsection{Contrastive learning for document representation}

Contrastive learning has emerged as a powerful paradigm for representation learning, initially popularized in computer vision through models such as SimCLR \cite{chen2020simple} and MoCo \cite{he2020momentum}. In natural language processing, contrastive methods have been applied to sentence and document embeddings, with models like SimCSE \cite{gao2021simcse,thirukovalluru2024sumcse} demonstrating strong performance on semantic similarity and retrieval tasks. In the scientific domain, contrastive learning has been adopted to enhance document embeddings based on citation links or topical proximity, as seen in SPECTER and SciNCL \cite{huang2024refcit2vec,liang2025beyond}. These approaches significantly improved retrieval quality by leveraging external supervision signals such as citation networks or semantic clusters.

Nevertheless, current contrastive learning frameworks face several limitations when applied to cross-disciplinary and multilingual scientific recommendation. They typically rely on single textual inputs, such as abstracts or titles, and therefore neglect heterogeneous components like objectives, methods, and results, which provide critical role-specific information. Furthermore, multilingual consistency is rarely incorporated, leading to suboptimal performance in cross-lingual retrieval scenarios. Finally, alignment between metadata and structured summaries is often overlooked, constraining the system’s ability to integrate multiple semantic perspectives of the same paper. Addressing these gaps, our study introduces role-specific projection heads and evidence-constrained OMRC summaries into a contrastive learning framework, enabling fine-grained, multilingual, and cross-disciplinary representations for scholarly recommendation.

\section{Methodology}
\label{methodology}

\subsection{Overall}

\begin{figure*}[t]
    \centering
    \includegraphics[width=0.98\linewidth]{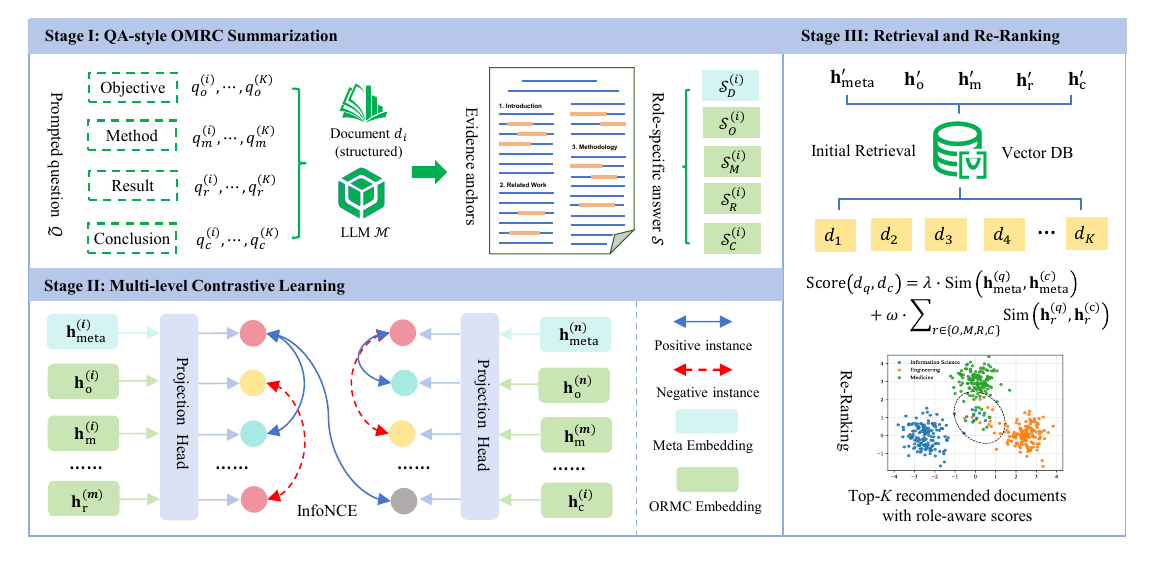}
\vspace{-.5cm}
\caption{Overall architecture of the proposed framework for cross-disciplinary and multilingual scholarly recommendation.}
\vspace{-.3cm}
\label{fig:Figure2}
\end{figure*}

Our framework comprises three sequential stages designed to enable cross-disciplinary and multilingual scholarly recommendation, as illustrated in Fig.~\ref{fig:Figure2}. 
In the first stage, a QA-style summarization module transforms each document into structured, evidence-grounded summaries following the OMRC schema. 
Subsequently, a multi-level contrastive learning process jointly optimizes document-level and role-level representations using a multilingual encoder and role-aware projection heads, ensuring both global semantic alignment and intra-document differentiation. 
In the final stage, the learned embeddings are employed for retrieval and re-ranking, where coarse-grained metadata matching is refined through fine-grained role-aware similarity computation. 
This hierarchical design bridges factual summarization and semantic retrieval, producing interpretable and robust representations for scholarly recommendation across domains and languages.

\begin{algorithm}[t]
\SetAlCapFnt{\footnotesize}
\caption{\footnotesize{OMRC-based Multi-level Retrieval and Re-ranking}}
\footnotesize
\label{alg:scispandet}
\KwIn{Corpus of documents $\mathcal{D}$; query document $d_q$; encoder $f_{\theta}$; role-aware projection heads $g_{\phi}^{(r)}$ ($r \in \{O,M,R,C\}$); balance coefficient $\lambda$; retrieval size $N$.}
\KwOut{Ranked list of relevant documents $\left\{d_c^{(1)}, \dots, d_c^{(N)}\right\}$.}

\ForEach{$\mathrm{document} \; d_i \in \mathcal{D} \;$}{
    Construct QA-style OMRC summaries $S_r^{(i)}$ grounded in textual evidence and align them with metadata summary $S_{\text{meta}}^{(i)}$; \\ 
    Encode metadata and OMRC summaries using $f_{\theta}$ and map them through $g_{\phi}^{(r)}$ to obtain structured embeddings $\mathbf{h}_{\text{meta}}^{(i)}, \mathbf{h}_r^{(i)}$.
}

After obtaining all structured embeddings $\left\{\mathbf{h}_{\text{meta}}^{(i)}, \mathbf{h}_r^{(i)}\right\}$, 
perform multi-level retrieval to identify semantically relevant documents.
\\[3pt]

\ForEach{$\mathrm{document} \; d_c \in \mathcal{D} \;$}{
   Compute global semantic similarity between the query document $d_q$ and the candidate $d_c$ using their metadata embeddings $\mathbf{h}_{\text{meta}}$;\\
    Record the similarity score for $d_c$.
}

Select top-$N$ documents with highest global similarity as $\mathcal{C}_N$;
\\[3pt]

\ForEach{ $\mathrm{candidate} \; d_c \in \mathcal{C}_N \;$}{
    Compute role-wise similarities between $\mathbf{h}_r^{(q)}$ and $\mathbf{h}_r^{(c)}$; \\
    Combine the global metadata similarity and the role-aware similarities using balance coefficient $\lambda$
    to obtain a composite score $\mathrm{Score}(d_q,d_c)$.
}

Sort $\mathcal{C}_N$ by $\mathrm{Score}(d_q,d_c)$ in descending order; \\
Return the top-$N$ ranked documents $\left\{d_c^{(1)}, \dots, d_c^{(N)}\right\}$ as the final recommendation list.
\end{algorithm}

\subsection{QA-style summarization}

To enable structured and interpretable representations of scientific papers, we design a QA-style OMRC summarization module that transforms each document into four role-specific summaries: \textit{Objective} (O), \textit{Method} (M), \textit{Result} (R), and \textit{Conclusion} (C). Unlike conventional abstractive summarization, which tends to produce generic or ungrounded content, our QA-style approach explicitly decomposes the summarization process into a set of evidence-constrained question–answering tasks, ensuring factual grounding, semantic completeness, and role consistency across disciplines and languages.

\noindent\textbf{Template-based Dynamic Augmentation.}
For each research role $r \in \left\{O, M, R, C\right\}$, we define a base question template $q_r^{\text{base}}$ that captures the semantic intent of that role (e.g., identifying objectives, describing methods, reporting results, and summarizing conclusions). These templates serve as stable semantic anchors, ensuring interpretability and structural consistency across documents.
To improve generalization across disciplines and languages, we extend each base template via dynamic paraphrased augmentation. A meta-prompting model generates a set of semantically equivalent variants $Q_r=\{q_r^{(1)}, q_r^{(2)}, \ldots, q_r^{(K)}\}$, introducing lexical and stylistic diversity while preserving meaning. Semantic fidelity is maintained through similarity-based filtering:
\begin{equation}
    Q_r^{*}=\left\{ q_r^{(k)} \,\middle|\, \mathrm{sim}\left(q_r^{(k)}, q_r^{\text{base}}\right)\ge \delta \right\},
\end{equation}
\noindent where cosine similarity $\mathrm{sim}(\cdot)$ ensures alignment with the base intent. In practice, we retain $K\in[5,8]$ variants per role with a threshold $\delta=0.85$. The resulting filtered set $Q_r^{*}$ constitutes a role-specific query pool that guides evidence-grounded summarization in the subsequent generation process. This hybrid mechanism combines the semantic stability of fixed templates with the contextual adaptability of dynamic generation, enabling robust QA-based summarization across heterogeneous academic domains.

\noindent\textbf{Evidence-Constrained QA Generation.}
Given the role-specific query pools $Q_r^{*}$, each document $d_i$ is processed through an evidence-constrained QA framework to produce factual, role-aware summaries. For each role $r\in\{O,M,R,C\}$, the summarization model $\mathcal{M}$ receives structured inputs that include section texts, figure/table captions, and explicit layout markers (e.g., \verb|<Section: Method>| or \verb|<Figure: 2>|). These structure-aware cues help the model locate relevant segments and maintain factual grounding during generation. Each query $q_r^{(k)}\in Q_r^{*}$ is paired with the document to form a QA instance:
\begin{equation}
    S_r^{(i)}=\mathcal{M}\left(d_i, q_r^{(k)}\right) \;\rightarrow\; \left(a_r^{(i,k)},\, E_r^{(i,k)}\right),
\end{equation}
\noindent where $a_r^{(i,k)}$ denotes the generated answer and $E_r^{(i,k)}\subset d_i$ represents the evidence anchors, including the text spans, figures, or tables cited in the response. During decoding, a relevance-weighted attention prior $w(e_j)$ is used to constrain generation toward evidentially aligned segments, mitigating factual drift.

To enhance stability, multiple paraphrased questions yield $K$ candidate answers per role. These candidates are embedded and clustered using cosine similarity, and the centroid representation is selected as the final summary:
\begin{equation}
    S_r^{(i)}=\mathrm{Aggregate} \left(\left\{\,a_r^{(i, k)}\,\right\}_{k=1}^{K}\right).
\end{equation}

This aggregation step filters redundancy and preserves semantic coherence, producing concise yet verifiable OMRC summaries.

\noindent\textbf{Structured Output Representation.}
The QA-based process yields a structured representation that captures both metadata and role-specific semantics. Formally,
\begin{equation}
    \mathcal{S}_i=\left\{S_{\text{meta}}^{(i)},\, S_O^{(i)},\, S_M^{(i)},\, S_R^{(i)},\, S_C^{(i)}\right\},
\end{equation}
\noindent where $S_{\text{meta}}^{(i)}$ encodes document metadata (title, abstract, and keywords), and $\left\{S_O^{(i)}, S_M^{(i)}, S_R^{(i)}, S_C^{(i)}\right\}$ are the role-specific summaries produced by the evidence-constrained QA procedure. Each $S_r^{(i)}$ provides a compact, factual, and interpretable view of the document’s objective, methodology, results, and conclusions. This structured output enables multi-perspective semantic alignment across documents and serves as the foundational input to the multi-level contrastive learning stage, ensuring that similarity is established not only at the document level but also across corresponding OMRC roles.

\subsection{Multi-level contrastive learning}

To effectively learn semantically aligned yet structurally discriminative representations, we design a multi-level contrastive learning framework based on the structured summaries $\mathcal{S}_i = \left\{ S_{\text{meta}}^{(i)}, S_O^{(i)}, S_M^{(i)}, S_R^{(i)}, S_C^{(i)} \right\}$ generated in the previous stage.
Each component $S_r^{(i)}$ is encoded by a multilingual encoder $f_\theta(\cdot)$ based on mBERT and projected through a role-aware transformation head $g_\phi^{(r)}(\cdot)$ to produce embeddings $\mathbf{h}_r^{(i)} = g_\phi^{(r)}\left(f_\theta(S_r^{(i)})\right)$.
This hierarchical design enables the model to capture both global document semantics and fine-grained role distinctions, forming the foundation for robust cross-disciplinary and multilingual retrieval.

\noindent\textbf{Document-level Contrastive Learning.}
At the document level, the model aligns documents with similar research themes while separating unrelated ones.
For each document $d_i$, its metadata summary $S_{\text{meta}}^{(i)}$ is encoded into an embedding $\mathbf{h}_{\text{meta}}^{(i)} = f_\theta\left(S_{\text{meta}}^{(i)}\right)$.
Pairs of documents belonging to the same research cluster serve as positive samples, while those from different clusters within the batch are treated as negatives.
The document-level contrastive objective is defined as:

\begin{equation}
\mathcal{L}_{\text{doc}}^{(i)}
= -\log \frac{\exp(\mathrm{sim}(\mathbf{h}_{\text{meta}}^{(i)}, \mathbf{h}_{\text{meta}}^{(i+)}) / \tau)}
{\sum_{j=1}^{N}\exp(\mathrm{sim}(\mathbf{h}_{\text{meta}}^{(i)}, \mathbf{h}_{\text{meta}}^{(j)}) / \tau)},
\end{equation}

\noindent where $\mathrm{sim}(\cdot)$ denotes cosine similarity, $\tau$ is a temperature hyperparameter, and $N$ is the batch size.
This objective encourages global semantic alignment among documents sharing similar research contexts, regardless of their disciplinary or linguistic differences.

\noindent \textbf{Role-level Contrastive Learning.}
To refine intra-document representations, we introduce a role-level contrastive objective that leverages the structured OMRC decomposition.
Each role-specific summary $S_r^{(i)}$ is encoded and projected to obtain $\mathbf{h}_r^{(i)}$.
Positive pairs consist of identical roles (e.g., $O$–$O$) across semantically related papers, whereas negatives are drawn from different roles or unrelated documents.
The loss is defined as:

\begin{equation}
\mathcal{L}_{\text{role}}^{(r,i)}
= -\log \frac{\exp(\mathrm{sim}(\mathbf{h}_r^{(i)}, \mathbf{h}_r^{+}) / \tau_r)}
{\sum_{j=1}^{N}\exp(\mathrm{sim}(\mathbf{h}_r^{(i)}, \mathbf{h}_r^{(j)}) / \tau_r)}.
\end{equation}

This mechanism enforces semantic coherence within each role while maintaining functional separation among $O$/$M$/$R$/$C$ components, ensuring that different discourse roles retain distinct representational spaces.

\noindent\textbf{Joint Optimization.}
The overall objective integrates both document-level and role-level contrastive signals:

\begin{equation}
\mathcal{L}_{\text{total}}
= \alpha \cdot \mathcal{L}_{\text{doc}}  + \beta \cdot \!\!\!\!\!\! \sum_{r\in \{O,M,R,C\}} \!\!\mathcal{L}_{\text{role}}^{(r)}.
\end{equation}

\noindent 
Here, $\alpha$ and $\beta$ balance global alignment and local role discrimination.
This joint optimization enables the encoder to learn a unified and structured semantic space that preserves document-level thematic proximity while maintaining clear distinctions across role-specific semantics.
As a result, the learned embeddings provide robust representations for cross-disciplinary retrieval and serve as the semantic foundation for the re-ranking stage.

\subsection{Retrieval and re-ranking}

Building upon the representations learned through multi-level contrastive learning, this stage performs retrieval and re-ranking to identify and prioritize the most relevant scientific papers for a given input document.

Each document $d_i$ is represented by its structured embedding set $\mathcal{H}_i = \left\{\mathbf{h}_{\text{meta}}^{(i)}, \mathbf{h}_O^{(i)}, \mathbf{h}_M^{(i)}, \mathbf{h}_R^{(i)}, \mathbf{h}_C^{(i)}\right\}$, derived from the encoder $f_{\theta}$ and role-aware projection heads $g_{\phi}^{(r)}$.
This structured representation enables both coarse-grained retrieval at the document level and fine-grained ranking at the role level.

\noindent\textbf{Coarse-grained Retrieval.}
In the coarse-grained stage, we perform semantic retrieval using the metadata embedding $\mathbf{h}_{\text{meta}}$, which encodes the overall context of each document.
Given an input document $d_q$ (query document), we compute its similarity with all candidate documents ${d_c}$ in the corpus using cosine similarity:

\vspace{-5pt}
\begin{equation}
\operatorname{Sim}_{\mathrm{doc}}\left(d_q, d_c\right)=\frac{\mathbf{h}_{\mathrm{meta}}^{(q)} \cdot \mathbf{h}_{\mathrm{meta}}^{(c)}}{\left\|\mathbf{h}_{\mathrm{meta}}^{(q)}\right\|\left\|\mathbf{h}_{\mathrm{meta}}^{(c)}\right\|}.
\end{equation}
\vspace{-2pt}

The system retrieves the top-$K$ most similar documents based on this similarity, where $K$ denotes the retrieval depth controlling the breadth of the initial search space.
From this set, the top-$N$ documents ($N \le K$) are retained as the candidate pool for the fine-grained re-ranking stage.
This hierarchical retrieval strategy ensures both efficiency and semantic coverage, enabling the subsequent role-aware re-ranking to focus on the most promising subset while maintaining high recall in large-scale scholarly corpora.

\noindent\textbf{Fine-grained Re-ranking.}
To refine initial retrieval results, role-aware re-ranking is performed using OMRC embeddings.
For each candidate $d_c$, a composite score is computed as a weighted sum of role-specific similarities:

\vspace{-5pt}
\begin{equation}
\begin{split}
\mathrm{Score}\left(d_q, d_c\right)
&= \lambda \cdot \operatorname{Sim}\left(\mathbf{h}_{\text{meta}}^{(q)}, \mathbf{h}_{\text{meta}}^{(c)}\right) \\
&\quad + \omega \cdot \!\!\!\!
\sum_{r \in \{O,M,R,C\}} \!\!\!
\operatorname{Sim}\left(\mathbf{h}_r^{(q)}, \mathbf{h}_r^{(c)}\right),
\end{split}
\end{equation}

\noindent
where $\lambda \in [0,1]$ balances the influence of global metadata and role-level similarities, and $\omega = (1-\lambda)/4$ normalizes the contribution of the OMRC components.
This formulation effectively captures cross-disciplinary relevance by aligning methodological or result-level semantics even when research objectives differ.
Candidate papers are then re-ranked according to $\mathrm{Score}(d_q, d_c)$, yielding a concise and interpretable recommendation list.

\section{Experiment and Analysis}

\begin{table*}[t]
\centering
\begin{threeparttable}
\caption{Comparison of recommendation performance on DBLP, S2ORC, and Sci-OMRC datasets.}
\label{tab:baseline_results}
\footnotesize 
\renewcommand{\arraystretch}{1.1}
\setlength{\tabcolsep}{6.9pt}
\rowcolors{1}{}{white}
\begin{tabular}{lcccccccccccc}
\toprule
\multirow{2}{*}{\textbf{Model}} & 
\multicolumn{4}{c}{\textbf{DBLP}} & 
\multicolumn{4}{c}{\textbf{S2ORC}} & 
\multicolumn{4}{c}{\textbf{Sci-OMRC}} \\
\cmidrule(lr){2-5}\cmidrule(lr){6-9}\cmidrule(lr){10-13}
 & Pre@10 & Rec@10 & ND@10 & MRR & Pre@10 & Rec@10 & ND@10 & MRR & Pre@10 & Rec@10 & ND@10 & MRR \\
\midrule
TF-IDF & 40.16 & 65.95 & 80.12 & 77.23 & 39.23 & 63.54 & 78.10 & 75.21 & 40.07 & 65.38 & 80.01 & 77.12 \\
BM25 & 42.58 & 68.31 & 81.53 & 78.41 & 41.20 & 65.72 & 79.58 & 76.34 & 42.35 & 68.02 & 81.28 & 78.23 \\
Doc2Vec & 48.85 & 72.46 & 82.84 & 79.55 & 46.72 & 69.16 & 81.26 & 77.89 & 48.56 & 72.13 & 82.56 & 79.27 \\
\noalign{\vskip 1pt}
\arrayrulecolor{black}\hdashline[3pt/3pt]\arrayrulecolor{black}
\noalign{\vskip 2pt}
Citeomatic & 55.24 & 76.20 & 84.55 & 81.42 & 53.12 & 73.80 & 83.64 & 80.02 & 55.01 & 76.08 & 84.13 & 81.15 \\
SciBERT & 56.12 & 77.15 & 85.01 & 81.93 & 54.05 & 74.26 & 84.17 & 80.43 & 55.92 & 77.01 & 84.85 & 80.64 \\
SPECTER & 60.08 & 79.36 & 86.82 & 83.48 & 57.94 & 77.13 & 86.05 & 82.28 & 61.02 & 79.65 & 86.97 & 83.84 \\
SciNCL & 61.76 & 80.57 & 87.56 & 84.10 & 58.56 & 78.06 & 86.71 & 83.07 & 62.24 & 80.76 & 87.62 & 84.26 \\
\noalign{\vskip 1pt}
\arrayrulecolor{black}\hdashline[3pt/3pt]\arrayrulecolor{black}
\noalign{\vskip 2pt}
\rowcolor[HTML]{ECF0FF}
\textbf{OMRC-MR} & \textbf{64.27} & \textbf{83.14} & \textbf{88.71} & \textbf{85.68} & \textbf{61.65} & \textbf{80.25} & \textbf{88.06} & \textbf{84.25} & \textbf{68.92} & \textbf{84.58} & \textbf{89.75} & \textbf{86.07} \\
\bottomrule
\end{tabular}
\vspace{-.2cm}
\end{threeparttable}
\end{table*}

\subsection{Experimental setup}
\noindent
\textbf{Datasets.}
We evaluate our approach on DBLP, S2ORC, and Sci-OMRC, covering both public bibliographic corpora and our curated structured dataset. The DBLP dataset \cite{sinha2015overview} is derived from the DBLP-Citation-Network V16\footnote{\url{https://open.aminer.cn/open/article?id=655db2202ab17a072284bc0c}} release, from which 10K papers published within the last five years are selected. The S2ORC corpus \cite{lo2020s2orc} is accessed via the official Semantic Scholar API\footnote{\url{https://api.semanticscholar.org/api-docs/recommendations}}, and 10K papers are sampled across multiple academic fields to enable cross-domain evaluation. The self-constructed Sci-OMRC dataset extends this setting by organizing 20K papers into structured scientific documents segmented into OMRC sections. Each record integrates the title, abstract, OMRC segments, and metadata, supporting fine-grained semantic representation for evidence-aware recommendation. All datasets are preprocessed to remove incomplete records and standardized into a unified input format containing the title, abstract, and metadata, while maintaining balanced distribution across disciplines. Detailed construction and preprocessing procedures are described in Appendix A.

\noindent
\textbf{Baseline methods.}
We compare our approach against a range of representative content-based recommendation baselines, all of which operate without user interaction data or citation graphs at inference time. These include: TF-IDF \cite{ni2021collaborative}, a lexical similarity baseline using cosine distance over sparse term-weighted vectors; BM25 \cite{rosa2021yes}, a probabilistic retrieval model that enhances TF-IDF with term saturation and length normalization; Doc2Vec \cite{zhengwei2022recommendation}, an unsupervised embedding model that learns dense representations by predicting contextual words; SciBERT \cite{beltagy2019scibert}, a transformer model pretrained on scientific corpora, from which we extract document embeddings via token averaging; Citeomatic \cite{bhagavatula2018content}, a hybrid citation recommender that combines TF-IDF and topic features for content-driven ranking; SPECTER \cite{cohan2020specter}, a SciBERT-based model trained with triplet loss over citation-informed paper triples, enabling content-only inference; and SciNCL \cite{ostendorff2022neighborhood}, a contrastive learning model that refines SPECTER via soft neighborhood sampling to better encode graded semantic similarity. All models are evaluated using the same structured input (title, abstract, metadata) under identical settings.

\noindent
\textbf{Implementation details.}
All experiments are conducted on a single NVIDIA A100 (80 GB) GPU. Embedding-based methods such as SciBERT, SPECTER, and SciNCL adopt the official checkpoints and are fine-tuned using the AdamW optimizer with a learning rate of 2e-5, batch size of 16, weight decay of 0.01, and up to 10 epochs with early stopping. Document embeddings are obtained by mean pooling over the final hidden states. TF-IDF and BM25 baselines are implemented with scikit-learn and Pyserini, and FAISS is used to construct a dense index for cosine similarity–based retrieval, where each test paper retrieves the top-10 most relevant documents within the same dataset. For the Sci-OMRC dataset containing full-text content, structured summaries are automatically generated through large language model calls using the Qwen3-Max API, ensuring consistent section segmentation and semantic completeness. All models share identical preprocessing pipelines, and results are averaged over three random seeds.

\subsection{Main results}

\begin{table}[t]
\centering
\begin{threeparttable}
\caption{Comparison of automated summarization models.}
\label{tab:doc-summ}
\footnotesize
\renewcommand{\arraystretch}{1.1}
\setlength{\tabcolsep}{6.5pt}
\begin{tabular}{lccccc}
\toprule
\textbf{Model} & \textbf{R-1} & \textbf{R-2} & \textbf{R-L} & \textbf{B-Score} & \textbf{G-Eval} \\
\midrule
TextRank \cite{ashari2017document}        & 35.95 & 12.68 & 26.43 & 56.86 & 50.54 \\
PGN \cite{anh2019abstractive}            & 38.87 & 14.16 & 28.15 & 61.83 & 56.38 \\
SciSummPip \cite{ju2020monash}     & 40.12 & 14.57 & 30.33 & 66.24 & 59.55 \\
Hie-BART \cite{akiyama2021hie}           & 40.31 & 14.73 & 30.48 & 67.16 & 59.86 \\
GPT-3.5-turbo \cite{achiam2023gpt}   & 47.26 & 22.75 & 35.57 & 76.05 & 73.43 \\
Qwen3-max \cite{yang2025qwen3}      & 49.13 & 23.56 & 37.68 & 79.32 & 74.84 \\
SciATS \cite{feng2025enhancing}         & 51.45 & 25.78 & 39.26 & 81.57 & 76.32 \\
\noalign{\vskip 1pt}
\arrayrulecolor{black}\hdashline[3pt/3pt]\arrayrulecolor{black}
\noalign{\vskip 2pt}
\rowcolor[HTML]{ECF0FF}
\textbf{QA-Style (Ours)} & \textbf{53.21} & \textbf{27.35} & \textbf{42.01} & \textbf{83.68} & \textbf{80.46} \\
\bottomrule
\end{tabular}
\vspace{.2cm}
\end{threeparttable}
\end{table}

\noindent
\textbf{Overall Recommendation Performance.}
Table~\ref{tab:baseline_results} presents the overall recommendation results on the DBLP, S2ORC, and Sci-OMRC datasets.
Across all benchmarks, the proposed OMRC-MR framework consistently outperforms traditional lexical, embedding-based, and transformer-based baselines, demonstrating its superior capacity to capture both semantic relevance and structural intent within scientific papers.

Classical lexical models (TF-IDF, BM25) rely heavily on surface-level token overlap, resulting in limited recall and weak robustness across domains.
Unsupervised embedding models such as Doc2Vec achieve moderate improvements by learning distributed contextual representations but remain inadequate for long and structurally complex texts.
Hybrid systems (Citeomatic) and transformer-based encoders (SciBERT, SPECTER, SciNCL) further enhance retrieval quality through domain-specific or citation-aware pretraining; however, these models still process documents as unstructured text, overlooking their inherent discourse organization.

In contrast, OMRC-MR integrates QA-style OMRC summarization, contrastive fine-tuning, and structure-aware re-ranking, yielding consistent and substantial improvements across all datasets.
On the most challenging Sci-OMRC corpus, OMRC-MR achieves Precision@10 = 68.92, Recall@10 = 84.58, and MRR = 86.07, surpassing the strongest baseline (SciNCL) by +7.2\%, +3.8\%, and +1.8\%, respectively.
Notably, for DBLP and S2ORC, where full-text content is unavailable and only metadata (\textit{titles}, \textit{abstracts}, and \textit{keywords}) are utilized, OMRC-MR still attains stable improvements over all baselines.
This confirms that the proposed framework generalizes well under restricted input settings and effectively leverages metadata-level semantics to enhance scholarly recommendation.

Overall, these results demonstrate that incorporating QA-style structural summarization and multi-level retrieval substantially improves both semantic coverage and contextual alignment.

\noindent
\textbf{Evaluation of Automated Summarization.}
To further assess the effectiveness of the QA-style summarization module integrated in OMRC-MR, we compare it with representative extractive, abstractive, and large language model–based summarization methods. As shown in Table~\ref{tab:doc-summ}, our QA-style approach achieves the best overall results across all metrics, outperforming both conventional neural summarizers (PGN, SciSummPip, Hie-BART) and recent LLM-based baselines (GPT-3.5-turbo, Qwen3-max). Notably, both SciATS and our method employ Qwen3-Max as the generation backbone to ensure a fair comparison, while differing in the summarization formulation: SciATS follows a single-pass structured summarization paradigm, whereas our approach adopts a QA-aligned OMRC framework that explicitly decomposes the summarization process into Objective, Method, Result, and Conclusion components.

Our QA-style method attains ROUGE-1 = 53.21, ROUGE-2 = 27.35, and ROUGE-L = 42.01, surpassing SciATS by +1.8, +1.6, and +2.7 points, respectively. The consistent gains across ROUGE, BERTScore, and G-Eval confirm that QA-aligned decomposition yields more coherent, complete, and factually faithful summaries. Furthermore, the improved factual consistency in G-Eval indicates that the evidence-constrained generation mechanism effectively grounds each QA response in verifiable textual evidence, ensuring semantic alignment with the source document. These results demonstrate that the improvement originates from the structured QA-based design rather than from differences in the underlying language model, validating the superiority and reliability of the QA-style summarization strategy.

\subsection{Ablation studies and analysis}

\noindent
\textbf{Module-level ablations.}
To examine the contribution of each module in the proposed OMRC-MR framework, we conduct a set of module-level ablation experiments, as summarized in Table~\ref{tab:ablation_result}.
Three core components are evaluated to examine their individual effects within the framework. The QA-style summarization (QA-Sum) module converts raw papers into structured OMRC representations, enhancing discourse coherence and semantic completeness. The multi-level contrastive learning (MCL) mechanism improves representation discrimination across metadata, section, and document levels. The structure-aware re-ranking module further optimizes retrieval by refining contextual similarity among top candidates.

When QA-Sum is removed, performance drops notably (Precision@10 = 61.47, MRR = 83.35), revealing that structured QA-style representations effectively capture the scientific discourse and improve semantic completeness.
Eliminating MCL leads to similar degradation (Precision@10 = 62.11, MRR = 83.72), demonstrating that contrastive optimization across hierarchical granularity substantially strengthens the embedding alignment among heterogeneous scholarly elements.
Disabling the re-ranking module results in a smaller but consistent performance decline (Precision@10 = 67.58, MRR = 85.41), confirming that the multi-stage retrieval refinement helps filter semantically weak candidates and promotes ranking precision.

Overall, the Full Model achieves the best results (Precision@10 = 68.92, Recall@10 = 84.58, MRR = 86.07), outperforming all ablated variants by clear margins.
These results verify that each module plays a complementary role: QA-based structural summarization provides high-level semantic organization, multi-level contrastive learning improves representation robustness, and the re-ranking stage ensures final retrieval precision.
Together, they contribute to the superior overall performance and robustness of the OMRC-MR framework.

\begin{table}[t]
\centering
\begin{threeparttable}
\caption{Module-level ablation results of OMRC-MR.}
\label{tab:ablation_result}
\footnotesize
\renewcommand{\arraystretch}{1.1}
\setlength{\tabcolsep}{9pt}
\begin{tabular}{lcccc}
\toprule
\textbf{Variant} & \textbf{Pre@10} & \textbf{Rec@10} & \textbf{ND@10} & \textbf{MRR} \\
\midrule
\quad w/o QA-Sum                 & 61.47 & 79.63 & 85.42 & 83.35 \\
\quad w/o MCL             & 62.11 & 80.05 & 85.83 & 83.72 \\
\quad w/o Re-rank          & 67.58 & 83.74 & 89.07 & 85.41 \\
\noalign{\vskip 1pt}
\arrayrulecolor{black}\hdashline[3pt/3pt]\arrayrulecolor{black}
\noalign{\vskip 2pt}
\rowcolor[HTML]{ECF0FF}
\textbf{Full Model}           & \textbf{68.92} & \textbf{84.58} & \textbf{89.75} & \textbf{86.07} \\
\bottomrule
\end{tabular}
\begin{tablenotes}
\footnotesize
\item Abbreviations: QA-Sum = QA-style summarization, MC = Multi-level contrastive learning.
\end{tablenotes}
\end{threeparttable}
\end{table}

\begin{figure}[t]
    \centering
    \includegraphics[width=1\linewidth]{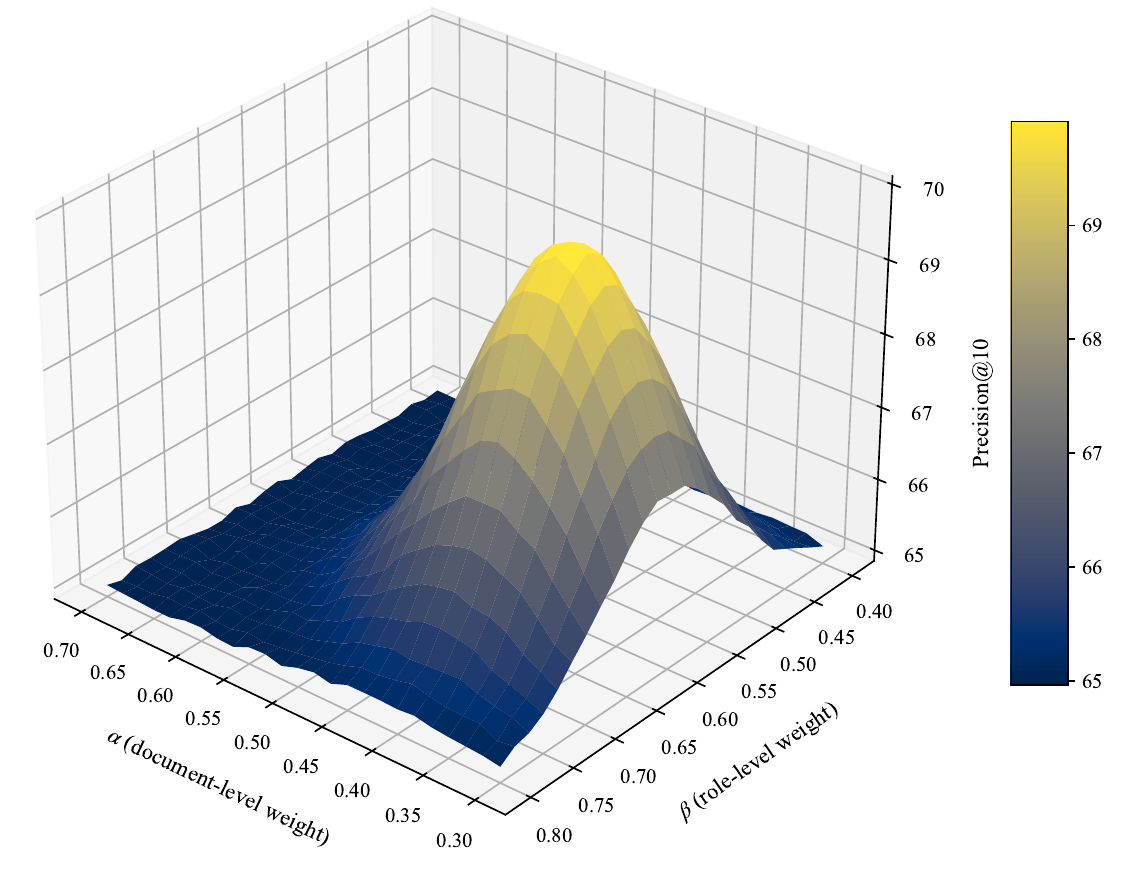}
\vspace{-.5cm}
\caption{
Performance variation under different weighting balances in joint contrastive learning, peaking at $\alpha$ = 0.4 and $\beta$ = 0.6.}
\label{fig:Figure3}
\end{figure}

\noindent
\textbf{Effect of weighting balance in joint contrastive learning.}
To analyze the impact of weighting balance in the joint contrastive learning objective, we systematically vary the coefficients $\alpha$ and $\beta$ that control the relative contributions of document-level and role-level objectives in Eq.~(7). The parameter $\alpha$ determines the strength of global semantic alignment across documents, while $\beta$ emphasizes local role-level discrimination among OMRC components.

As illustrated in Fig.~\ref{fig:Figure3}, the model’s performance, measured by Precision@10, forms a smooth surface with a clear performance peak around $\alpha = 0.4$ and $\beta = 0.6$. This pattern indicates that assigning higher weight to role-level contrastive learning leads to more effective embedding representations. When $\alpha$ dominates, the optimization overemphasizes document-level similarity, causing the model to capture broad thematic relations while neglecting finer role distinctions. In contrast, when $\beta$ becomes too large, the model loses overall semantic coherence, resulting in fragmented representations and reduced retrieval precision.

The optimal balance at $\alpha = 0.4$ and $\beta = 0.6$ demonstrates that emphasizing role-level contrastive objectives slightly more than document-level ones yields the best trade-off between global semantic consistency and discourse-level discrimination. This configuration allows the encoder to construct a structured yet coherent semantic space, improving the model’s interpretability and robustness in cross-disciplinary recommendation tasks.

\noindent
\textbf{Effect of structured summarization, role granularity, and evidence constraints.}
To comprehensively assess the internal mechanisms of the proposed QA-style OMRC summarization, we analyze how the summarization paradigm, role granularity, and evidence constraints collectively influence recommendation performance under identical encoder, training, and re-ranking settings. The study compares single-pass and QA-style generation to evaluate the impact of discourse-guided summarization, examines the necessity of complete four-role OMRC decomposition for capturing rhetorical structure, and investigates how explicit structural markers and evidence anchors enhance factual grounding and semantic alignment.
Together, these analyses aim to determine how structural decomposition, discourse granularity, and evidence grounding contribute to the effectiveness of the proposed framework.

\begin{table}[t]
\centering
\begin{threeparttable}
\caption{Performance under different summarization paradigms, role granularities, and evidence settings on the Sci-OMRC dataset (encoder, training, and re-ranking fixed).}
\label{tab:summ_role}
\footnotesize
\renewcommand{\arraystretch}{1.1}
\setlength{\tabcolsep}{8pt}
\begin{tabular}{lcccc}
\toprule
\textbf{Configuration} & \textbf{Pre@10} & \textbf{Rec@10} & \textbf{ND@10} & \textbf{MRR} \\

\midrule
\rowcolor[HTML]{F0F0F0}
\multicolumn{5}{c}{\textit{Summarization Paradigm Analysis}} \\
\midrule

Single-pass (LLM) & 63.46 & 81.02 & 87.25 & 84.02 \\
QA-style (w/o aug.) & 66.75 & 82.87 & 88.34 & 85.14 \\
\noalign{\vskip 1pt}
\arrayrulecolor{black}\hdashline[3pt/3pt]\arrayrulecolor{black}
\noalign{\vskip 2pt}
\rowcolor[HTML]{ECF0FF}
QA-style (w/ aug.) & \textbf{68.92} & \textbf{84.58} & \textbf{89.75} & \textbf{86.07} \\

\midrule
\rowcolor[HTML]{F0F0F0}
\multicolumn{5}{c}{\textit{Role granularity Analysis}} \\
\midrule

O+M & 64.28 & 81.16 & 87.34 & 83.92 \\
O+M+R & 67.04 & 83.13 & 88.69 & 85.18 \\
\noalign{\vskip 1pt}
\arrayrulecolor{black}\hdashline[3pt/3pt]\arrayrulecolor{black}
\noalign{\vskip 2pt}
\rowcolor[HTML]{ECF0FF}
O+M+R+C (OMRC) & \textbf{68.92} & \textbf{84.58} & \textbf{89.75} & \textbf{86.07} \\

\midrule
\rowcolor[HTML]{F0F0F0}
\multicolumn{5}{c}{\textit{Evidence Constraints}} \\
\midrule

Evidence-free QA & 66.03 & 82.11 & 88.07 & 84.72 \\
\noalign{\vskip 1pt}
\arrayrulecolor{black}\hdashline[3pt/3pt]\arrayrulecolor{black}
\noalign{\vskip 2pt}
\rowcolor[HTML]{ECF0FF}
Evidence-constrained & \textbf{68.92} & \textbf{84.58} & \textbf{89.75} & \textbf{86.07} \\

\bottomrule
\end{tabular}
\end{threeparttable}
\end{table}

As shown in Table \ref{tab:summ_role}, replacing single-pass summaries with QA-style structured summaries markedly improves recommendation performance, yielding gains of +5.46 Precision@10 and +2.05 MRR. This demonstrates that role-consistent and evidence-grounded summarization provides a stronger semantic basis for contrastive learning by preserving the discourse organization of scientific papers. Moreover, template-based dynamic augmentation further enhances cross-disciplinary robustness, as paraphrased question templates enrich semantic diversity across research domains and languages.

The role granularity analysis further confirms the necessity of the complete four-role decomposition. Adding the \textit{Result} (R) and \textit{Conclusion} (C) roles yields consistent improvements across all metrics (+2.64 Precision@10, +1.25 MRR), as these roles capture outcome-oriented and interpretive semantics that generalize effectively across disciplines. This indicates that the four-role OMRC schema is not heuristic but principled, modeling the full rhetorical flow of scholarly communication.

Removing explicit structural cues such as section markers and figure or table anchors results in performance declines of –2.89 Precision@10 and –1.35 MRR, indicating weaker factual grounding and reduced contextual coherence. These results confirm that the performance gains of OMRC-MR arise from effectively leveraging the intrinsic structural information of scientific documents, rather than from superficial prompt modifications.

\begin{figure}[t]
    \centering
    \includegraphics[width=0.9\linewidth]{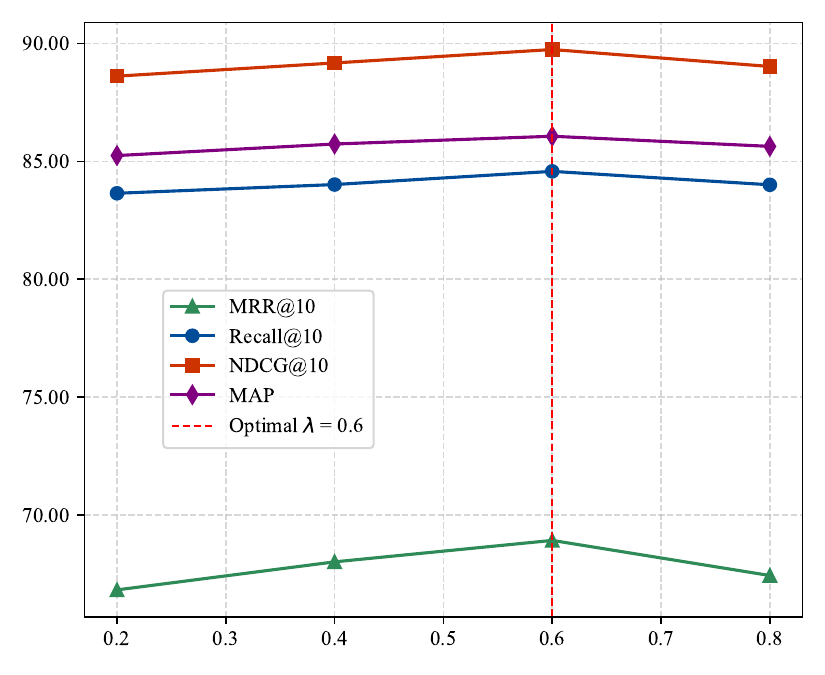}
\vspace{-.4cm}
\caption{
Retrieval performance under different re-ranking weights ($\lambda$), showing peak precision and ranking quality at $\lambda$ = 0.6, where metadata and role-aware similarities are optimally balanced.}
\label{fig:Figure4}
\end{figure}

\noindent
\textbf{Effect of re-ranking strategy on retrieval performance.}
To examine the impact of the re-ranking mechanism, we vary the weighting coefficient $\lambda$ in Eq.~(11), which controls the balance between metadata-based similarity and role-aware similarity across OMRC components. As shown in Fig.~\ref{fig:Figure4}, all evaluation metrics exhibit a consistent trend with the optimal performance observed at $\lambda = 0.6$.

Among the metrics, Precision@10 shows the largest variation, increasing from 66.82 to 68.92 as $\lambda$ grows from 0.2 to 0.6 and then slightly declining to 67.43 at 0.8. This indicates that the re-ranking weight primarily influences top-rank precision, effectively refining the order of the most relevant papers. NDCG@10 and MAP follow similar but smoother trends, reflecting stable improvements in overall ranking quality and first-hit accuracy, while Recall@10 remains relatively unchanged.

These results demonstrate that emphasizing role-aware similarity up to a moderate level ($\lambda$ = 0.6) allows the system to capture both global thematic alignment and local discourse correspondence. Overemphasizing either metadata or role cues, however, leads to a decline in ranking precision, confirming that a balanced re-ranking strategy is critical for achieving both accuracy and interpretability in scholarly recommendation.

\begin{figure}[t]
    \centering
    \includegraphics[width=1\linewidth]{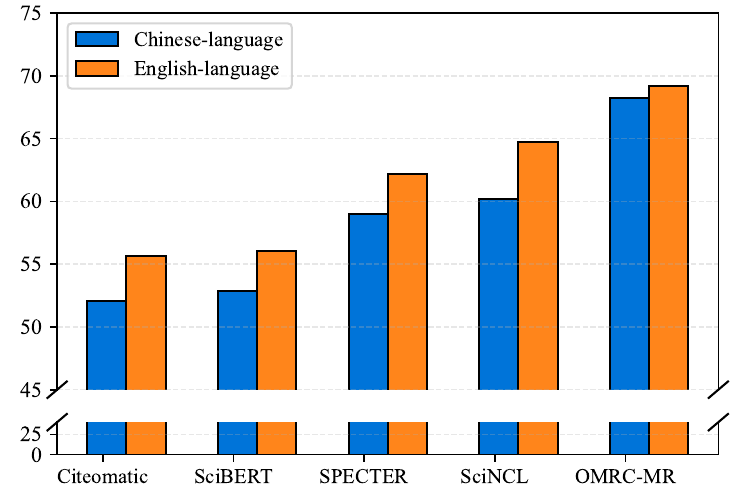}
\vspace{-.5cm}
\caption{
Cross-lingual Precision@10 comparison on the Sci-OMRC dataset.}
\label{fig:Figure5}
\end{figure}

\noindent
\textbf{Effect of language variation on model performance.}
To further assess the robustness of different models across languages, we compare representative methods including Citeomatic, SciBERT, SPECTER, SciNCL, and OMRC-MR on the English and Chinese subsets of the Sci-OMRC dataset using Precision@10 as the evaluation metric. As illustrated in Fig.~\ref{fig:Figure5}, all models exhibit consistent cross-lingual performance trends, with English scores generally exceeding Chinese results by about 2–3 percentage points, reflecting the broader pretraining coverage and vocabulary diversity in English corpora. For clarity, this experiment focuses on recent embedding-based methods and does not include traditional lexical baselines.

Among the compared models, SciNCL and SPECTER perform better than SciBERT and Citeomatic, highlighting the advantage of contrastive objectives in semantic alignment. OMRC-MR achieves the highest Precision@10 in both languages (69.15\% for English and 68.24\% for Chinese) and shows the smallest cross-lingual gap, indicating strong robustness and stable semantic representation across linguistic domains. These results confirm that the proposed framework effectively preserves semantic integrity and ranking precision in multilingual scholarly recommendation tasks.

\noindent
\textbf{Effect of disciplinary variation on model performance.}
To evaluate the robustness of different models across academic domains, we analyze their performance on eight major disciplines within the Sci-OMRC dataset, including Fundamental \textit{Science}, \textit{Medicine}, \textit{Agricultural}, \textit{Engineering}, \textit{Social}, \textit{Economics}, \textit{Philosophy}, and \textit{Information}. Precision@10 is used as the primary metric to ensure consistent evaluation across models and domains.

\begin{figure}[t]
    \centering
    \includegraphics[width=1\linewidth]{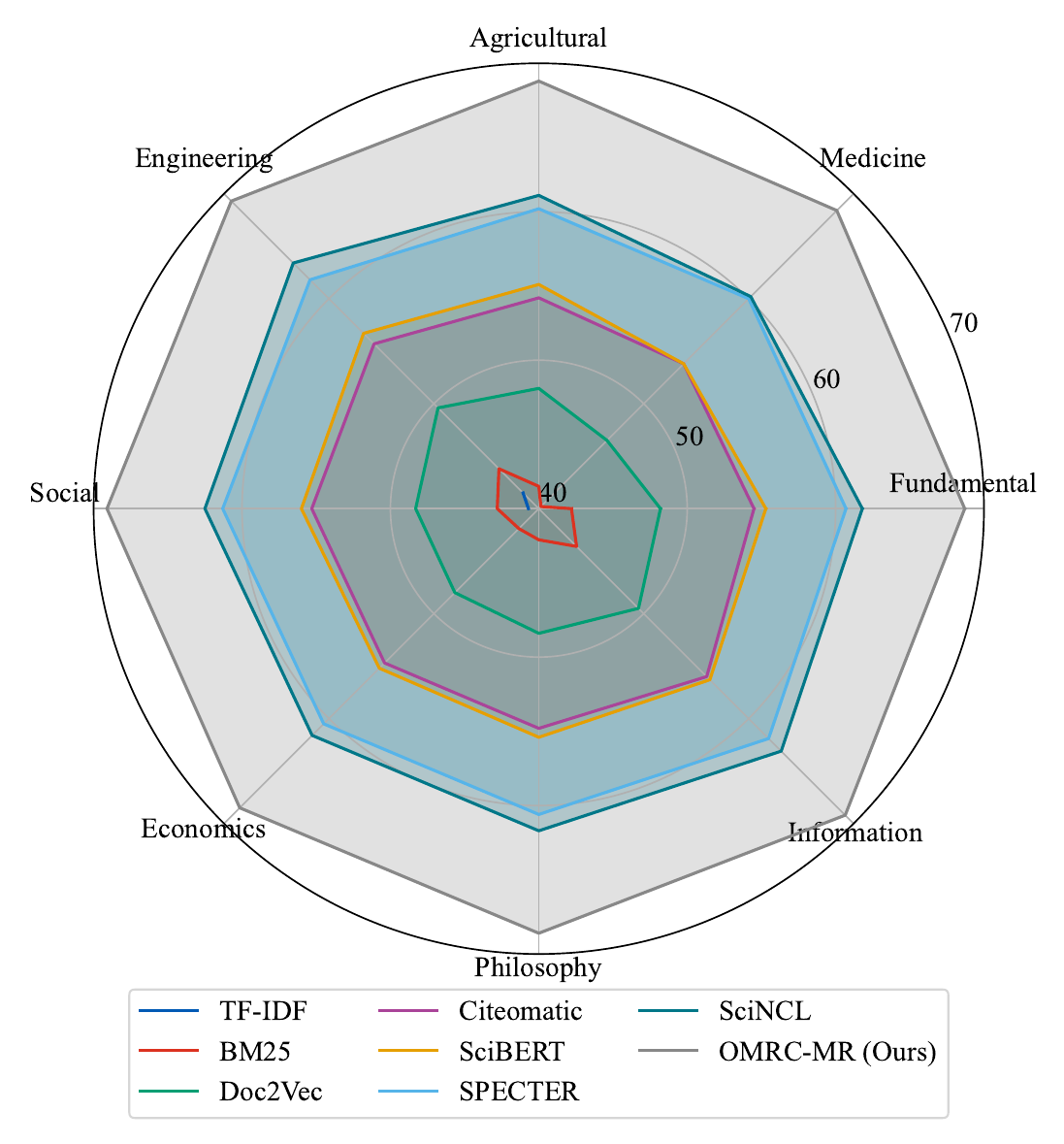}
\vspace{-.5cm}
\caption{
Precision@10 across multiple disciplines in the Sci-OMRC dataset, showing OMRC-MR with the most consistent top performance.}
\label{fig:Figure6}
\end{figure}

As illustrated in Fig.~\ref{fig:Figure6}, the performance trends remain consistent across all disciplines, reflecting the generalization ability of embedding-based recommendation models. TF-IDF, BM25, and Doc2Vec show limited variation across disciplines, indicating that lexical similarity alone captures only surface-level relevance. Representation learning methods, such as SciBERT and SPECTER, achieve steady improvements, particularly in \textit{Engineering} and \textit{Information}, where textual structures are more standardized and terminology is well-represented in pretraining corpora. SciNCL further enhances cross-domain discrimination through contrastive learning, yielding more stable precision across heterogeneous fields.

OMRC-MR achieves the highest Precision@10 in all disciplines, with scores ranging from 68.4\% to 69.3\%. The small variance (within 1.0 point) demonstrates strong domain invariance and adaptability. This stability is attributed to the model’s role-aware contrastive learning and re-ranking mechanisms, which effectively align global thematic relevance with discipline-specific discourse structures. The results confirm that OMRC-MR generalizes well across diverse scientific domains without requiring field-specific tuning, underscoring its robustness for cross-disciplinary scholarly recommendation.

\begin{figure}[t]
    \centering
    \includegraphics[width=1\linewidth]{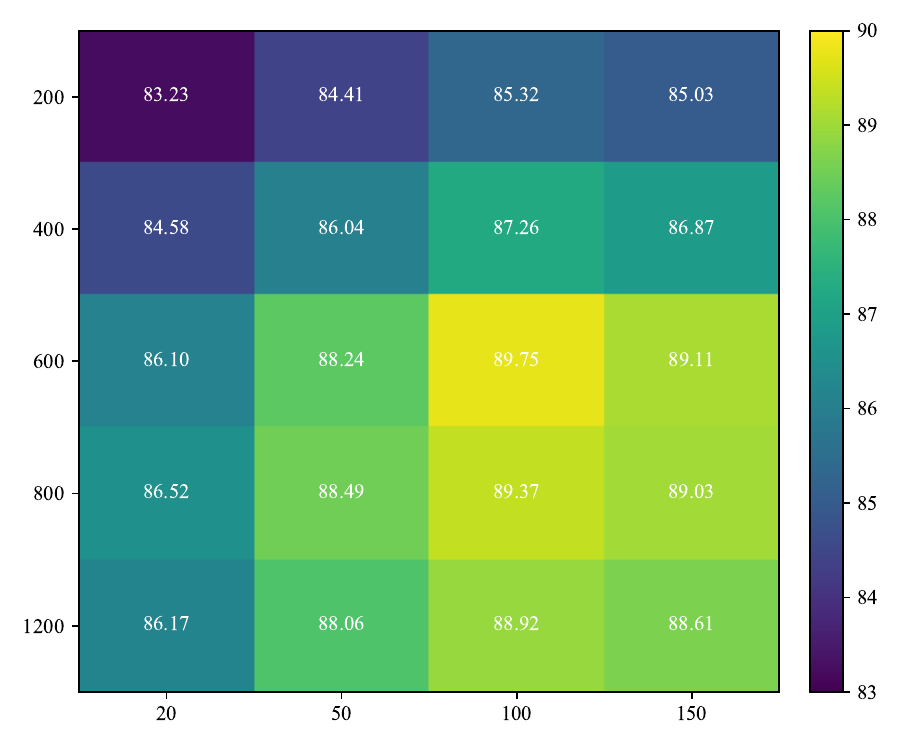}
\vspace{-.5cm}
\caption{
Effect of retrieval depth ($K$) and candidate pool size ($N$) on NDCG@10, with optimal configuration at $K$ = 600 and $N$ = 100.}
\label{fig:Figure7}
\end{figure}

\noindent
\textbf{Effect of retrieval depth and candidate pool size.}
To further investigate the retrieval sensitivity of the proposed framework, we analyze how the retrieval depth ($K$) and candidate pool size ($N$) affect ranking performance, as measured by NDCG@10. The results, visualized in Fig.~\ref{fig:Figure7}, reveal a clear two-dimensional trend: model performance consistently improves with increasing K and N until reaching an optimal configuration, beyond which marginal gains diminish or slightly reverse.

Specifically, NDCG@10 rises steadily from 83.23 at $K$ = 200, $N$ = 20 to a peak of 89.75 at $K$ = 600, $N$ = 100, indicating that a moderate retrieval depth and balanced candidate scope yield the best trade-off between coverage and precision. Further enlarging $K$ or $N$ beyond these thresholds introduces redundant or noisy candidates, causing a mild decline in ranking quality, as reflected by the decrease to 88.61 when $K$ = 1200 and $N$ = 150.

These observations suggest that excessive expansion of the retrieval pool may degrade precision by incorporating semantically distant papers, while insufficient depth limits contextual recall. Therefore, maintaining a moderate retrieval depth ($K$ = 600) and candidate pool size ($N$ = 100) ensures an optimal balance between efficiency and ranking fidelity, validating the scalability and robustness of our hierarchical retrieval design.

\section{Conclusion}

This study introduced OMRC-MR, a discourse-aware and fully content-based framework for scientific paper recommendation. By integrating QA-style OMRC summarization, multi-level contrastive learning, and structure-aware re-ranking, the framework explicitly models the rhetorical structure of scientific discourse and aligns multi-granular semantic representations across metadata, section, and document levels. This design enables interpretable and fine-grained recommendation while remaining entirely independent of citation or interaction data, thereby ensuring scalability and privacy preservation.

Extensive experiments on DBLP, S2ORC, and the newly constructed Sci-OMRC dataset demonstrate that OMRC-MR consistently outperforms state-of-the-art baselines, highlighting the critical role of discourse structure and hierarchical semantic alignment in improving retrieval precision and generalization. Future extensions will explore enhanced discourse decomposition beyond OMRC, adaptive contrastive objectives for heterogeneous document elements, and cross-domain transfer strategies to further strengthen the framework’s capability for scientific knowledge understanding and synthesis.

\bibliographystyle{elsarticle-num} 
\bibliography{refer}

\end{document}